\documentclass[10pt]{article}
\usepackage[OE]{express}
\pdfoutput=1

\newcommand{\enquote}[1]{``#1''}

\begin{document}

\title{Generation of highly pure Schr{\"o}dinger's cat states and real-time quadrature measurements via optical filtering}

\author{Warit Asavanant,\authormark{1,3} Kota Nakashima,\authormark{1} Yu Shiozawa,\authormark{1} Jun-ichi Yoshikawa,\authormark{1,2} and Akira Furusawa\authormark{1,4}}

\address{\authormark{1}Department of Applied Physics, School of Engineering, The University of Tokyo, 7-3-1 Hongo, Bunkyo-ku, Tokyo, 113-8656, Japan}

\address{\authormark{2}Quantum-Phase Electronics Center, School of Engineering, The University of Tokyo, 7-3-1 Hongo, Bunkyo-ku, Tokyo, 113-8656, Japan}
\email{\authormark{3}warit@alice.t.u-tokyo.ac.jp} 
\email{\authormark{4}akiraf@ap.t.u-tokyo.ac.jp}



\begin{abstract*}
Until now, Schr{\"o}dinger's cat states are generated by subtracting single photons from the whole bandwidth of squeezed vacua. However, it was pointed out recently that the achievable purities are limited in such method (J. Yoshikawa, W. Asavanant, and A. Furusawa, arXiv:1707.08146 [quant-ph] (2017)). In this paper, we used our new photon subtraction method with a narrowband filtering cavity and generated a highly pure Schr{\"o}dinger's cat state with the value of $-0.184$ at the origin of the Wigner function. To our knowledge, this is the highest value ever reported without any loss corrections. The temporal mode also becomes exponentially rising in our method, which allows us to make a real-time quadrature measurement on Schr{\"o}dinger's cat states, and we obtained the value of $-0.162$ at the origin of the Wigner function.
\end{abstract*}



\section{Introduction}
Generation of highly nonclassical states with high purity and fidelity is challenging but necessary for quantum information processing. One of the states with such high nonclassicality is a quantum superposition of macroscopically distinguishable states; so-called Schr{\"o}dinger's cat states with the name taken from the famous Schr{\"o}dinger's cat paradox. The optical Schr{\"o}dinger's cat states refer to the superposition of coherent states with phase difference of $\pi$ and can be written as $\vert\Psi_{\textrm{cat},\pm}\rangle\propto\vert\alpha\rangle\pm\vert-\alpha\rangle$, where $\vert\Psi_{\textrm{cat},\pm}\rangle$ corresponds to plus and minus Schr{\"o}dinger's cat states, respectively. Cat states do not only pose interest in fundamental quantum physics, but also possess many potential applications, such as quantum computation \cite{Jeong2002b,Ralph2003b,Lund2008b,Neergaard-Nielsen2010d}, entanglement distribution and quantum key distribution \cite{Karimipour2002a,Brask2010d,Sangouard2010a}, and quantum metrology \cite{Gilchrist2004a}. In optical continuous-varible (CV) quantum information processing, the measurement-based quantum computation (MBQC) is currently the most promising method in terms of operation implementation and scalability \cite{Gottesman1999,Raussendorf2001}. 
Therefore, the generation of high fidelity and large amplitude $\alpha$ cat states and the combination of such states with MBQC will take us a step closer to the realization of quantum information processing.

\begin{figure}[t!]
\centering\includegraphics[width=0.8\textwidth]{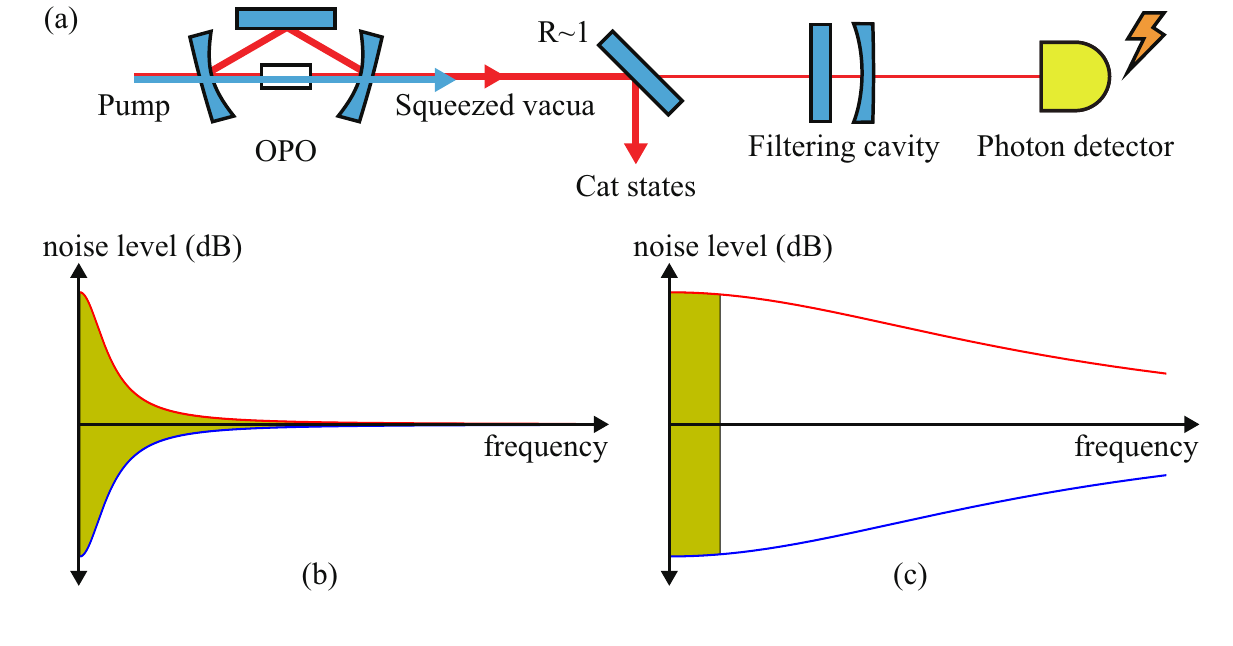}
\caption{Schematic diagram of photon subtraction and comparison of relation of squeezed spectra and frequency bandwidth of subtracted photon between previous demonstrations and our optical filtering method. (a) Schematic diagram of photon subtraction. (b) Previous demonstrations.  (c) our method. Red line: anti-squeezing level. Blue line: squeezing level. Area painted with yellow: frequency bandwidth of photon subtracted from squeezed state. }
\label{fig:ourmethod}
\end{figure}

The first proposition for the generation of cat states was via the Kerr effect in a nonlinear medium \cite{Yurke1986a}. However, the Kerr effect is usually weak. Instead, the generation by photon subtraction \cite{Dakna1997a,Lund2004} is widely used due to its simplicity in the implementation. In photon subtraction, a single photon is tapped from a squeezed vacuum, and the generated state is heralded by the detection of subtracted single photon (Fig. \ref{fig:ourmethod}(a)). One can generate states with high fidelity to single photon states, cat states, and squeezed single photon states, depending on the initial squeezing level. The cat states generated by this method are usually limited to the case where the amplitude is small, i.e. $\vert\alpha\vert^{2}\approx 1$. The generation of cat states using photon subtraction has already been demonstrated in both pulsed regime \cite{Ourjoumtsev2006a} and continuous-wave (CW) regime \cite{Neergaard-Nielsen2006a,Wakui2007a}. Two photons subtraction \cite{Takahashi2008a} and three photons subtraction \cite{Gerrits2010a} have also been experimentally demonstrated. Moreover, more complex quantum states, such as parity qubits based on cat states \cite{Neergaard-Nielsen2010d} and entanglement between single-rail qubits and cat states \cite{Morin2014b} have also been generated by method which involves photon subtraction. 

However, in the recent work by Yoshikawa {\it et al.} \cite{Junichi2017}, it was shown that the frequency bandwidth of the subtracted photons, which has not been given much attention, affects the purity of the generated cat states. In the previous demonstration of generation of cat states \cite{Ourjoumtsev2006a,Neergaard-Nielsen2006a,Wakui2007a}, where they used the whole frequency bandwidth of squeezed vacua for photon subtraction (Fig. \ref{fig:ourmethod}(b)), the achievable purities are limited by the inherent impurities of the initial squeezed vacua. This is due to the fact that when the cat states are heralded by the detection of subtracted single photons, the heralded states exist in a wave packet. In frequency domain, the squeezed vacuum is pure for all frequency. However, when we consider a squeezed state in a wave packet in time domain, the squeezed vacua become impure, and this impurity affects the subsequently generated cat states. Also, the inherent impurity in squeezing level due to the shape of the wave packets limits the squeezing level before the photon subtraction. In typical photon subtraction, where the temporal mode is double-side exponential, such impurity is calculated to be equivalent to 10\% loss \cite{Junichi2017}. Therefore, roughly speaking, the conventional photon subtraction can be performed on squeezing vauca with squeezing level of atmost about 10 dB. On the other hand, our method does not have such limitation.

As mentioned before, MBQC is currently the most realistic method for the realization of practical quantum information processing. In MBQC, the input states are entangled with the resource state called cluster state, and the operations are carried out by measurement on each modes of cluster state. When the modes are measured, the measurement results need to be fed forward to the next step of the operations. Linear feed-forward operations for each measurements can be postponed and performed after all measurements have been finished \cite{Gu2009}. On the other hand, nonlinear feed-forward operations cannot be postponed and have to be done before the next operation step. This means that the acquisitions of the quadrature values must be performed in real time in order to realize efficient non-linear feed-forward operations. Therefore, real-time quadrature measurement is an important piece in the universal MBQC since such non-linear feed-forward operations are  necessary for non-Gaussian operation \cite{Miyata2016}, which will allow us to perform quantum information processing that surpass classical computation \cite{Bartlett2002,Gottesman2001}. 

Real-time quadrature measurement of single photon states has already been demonstrated by generating single photon states with an exponentially rising temporal mode \cite{Ogawa2016}. However, in their method, non-degenerated and asymmetric optical parametric oscillator (OPO) is used for the temporal mode shaping. Therefore, their method is not applicable to quantum states that need degenerated OPO in the generation, such as cat states. Another method of temporal mode shaping using optical cavities has already succeeded in the case of single photon states \cite{Qin2015,Srivathsan2014}. However, there has not been any demonstration of the temporal mode shaping of the quantum states with multiple photons or with phase information.
 
In this paper, we use narrowband filtering cavity in the subtracted photon path to limit the frequencies of the subtracted photons to where the squeezing level is almost constant (Fig. \ref{fig:ourmethod}(c)). By doing so, we can generate pure cat states \cite{Junichi2017}. In the previous demonstration of cat states \cite{Ourjoumtsev2006a,Neergaard-Nielsen2006a,Wakui2007a}, filtering cavities are already used for filtering out non-degenerated photon pairs. However, the frequency bandwidth of the filtering cavity is much wider than that of OPO, so that raw photon correlation of OPO can be utilized. In the time domain, spectral filtering of subtracted photons is equivalent to the temporal mode shaping of the cat states. In previous demonstrations, the temporal mode of cat states become double-sided exponential due to photon correlation of OPOs. On the other hand, in our method, the temporal mode becomes exponentially rising due to the frequency response of the narrowband filtering cavity. By utilizing the exponentially rising temporal mode, we can realize the real-time measurement of cat states. The realization of real-time measurement of cat states will facilitate the usage of cat states in MBQC, thus expanding the potential of MBQC.

\section{Experimental setup}
\begin{figure}[t]
\centering
\includegraphics[width=0.7\textwidth]{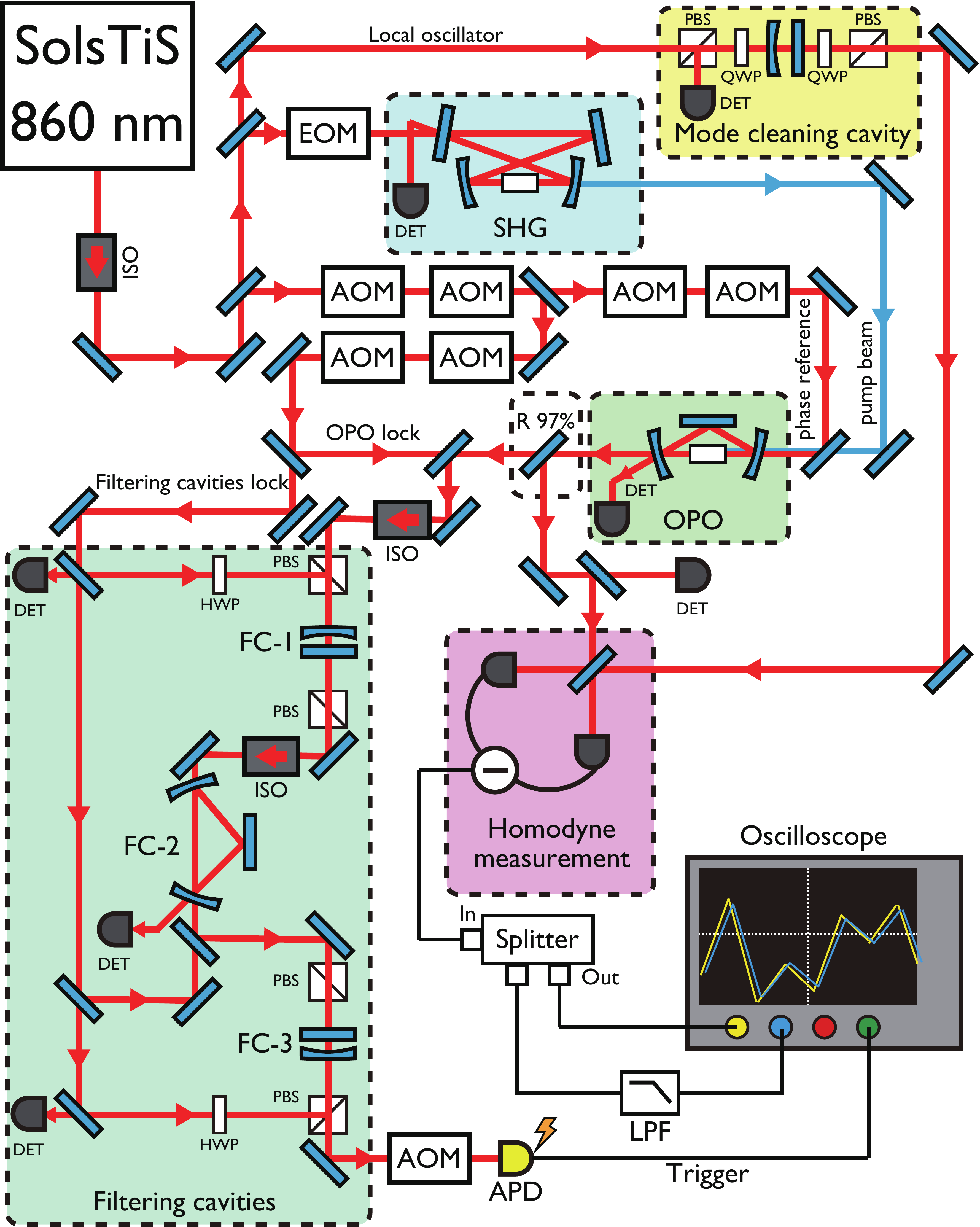}
\caption{Schematic diagram of the experiment setup. Only important optical components are included in the figure. Piezoelectric components and electrical channels for feedback and control are omitted. Red lines denote 860 nm laser beam, while blue lines denote 430 nm laser beam. SHG: Second harmonic generator. ISO: Isolator. HWP: Half-wave plate. QWP: Quarter-wave plate. PBS: Polarization beamsplitter. EOM: Electro-optic modulator. AOM: Acousto-optic modulator. APD: Avalanche photodiode. FC: Filtering cavity. LPF: Low-pass filter. DET: Detector. Detectors shown in the figure are for detecting error signal used in cavities locking and such.}
\label{fig:experiment}
\end{figure}
We used a broadband OPO and a narrowband filtering cavity to demonstrate the generation of highly pure cat states by optical filtering (Fig. \ref{fig:ourmethod}(c)). In \cite{Lee2011b}, cat state was generated with an OPO with bandwidth of approximately 10 MHz. Therefore, in order to demonstrate photon subtraction with narrow bandwidth using the OPO with the same specification, we have to use filtering cavity which has much narrower bandwidth (for example, 1 MHz) . In that case, the generation rate will drop drastically which is not favorable experimentally, and the effect of low frequency noises such as laser noise might also become more apparent. To avoid these problems, we made a broadband OPO and made a filtering cavity whose bandwidth is narrow compare to this OPO (Fig. \ref{fig:ourmethod}(c)).
 
The schematic diagram of the experiment setup is shown in Fig. \ref{fig:experiment}. A CW Ti:Sapphire laser (M-squared, Sols:TiS) whose wavelength is 860 nm is used as a light source for this experiment. A 430 nm CW pump beam for the OPO is produced from a bow-tie shaped second harmonic generator. A rectangular shaped reference cavity (not shown in Fig. \ref{fig:experiment}) between the second harmonic generator and the OPO is provided for matching the transversal modes of the pump beam and the OPO. The OPO used in this experiment is a triangle-shaped cavity with the same design as in \cite{Serikawa2016} with the linewidth of 130 MHz and a 10 mm long periodically poled KTiOPO\textsubscript{4} (PPKTP) crystal is placed inside the OPO and used as a nonlinear medium.

A beamsplitter with reflectivity of $R=0.97$ is used for tapping a single photon from squeezed vacuum. There are three filtering cavities (FC-1, FC-2, FC-3) on the subtracted photon path. FC-1 and FC-3 are Fabry-P{\`e}rot cavities with large free spectral range (FSR) and acts as frequency filters that filter out unwanted non-degenerated photon pairs. On the other hand, FC-2 is a triangle-shaped narrowband cavity with full-width half-maximum (FWHM) much narrower than the OPO. This cavity is used to limit the frequency bandwidth of the subtracted photons. The detailed parameters of all the filtering cavities are shown in Table \ref{tab:filterpara}. We put an isolator between FC-1 and FC-2 to prevent coupling between FC-1, FC-2 and FC-3. We also put another isolator between the OPO and FC-1 to prevent lock beams of filtering cavities from reaching the OPO.
\begin{table}[t!]
\caption{Parameters of the OPO and filtering cavities}
\centering
\begin{tabular}{*{1}{l}*{5}{c}}
\hline
&Shape&Length (mm)&Reflectivity&FSR (GHz)&FWHM (MHz)\\
\hline
OPO&Triangle&46.3&HR, HR, 0.86&5.32&130\\
FC-1&Fabry-P{\`e}rot&1.99&0.994, 0.994&75.4&136\\
FC-2&Triangle&103&0.98, HR, 0.98&2.91&18.7\\
FC-3&Fabry-P{\`e}rot&2.89&0.994, 0.994&51.9&94.0\\
\hline
\end{tabular}
\label{tab:filterpara}
\end{table}
\begin{figure}[t!]
\centering\includegraphics[width=0.5\textwidth]{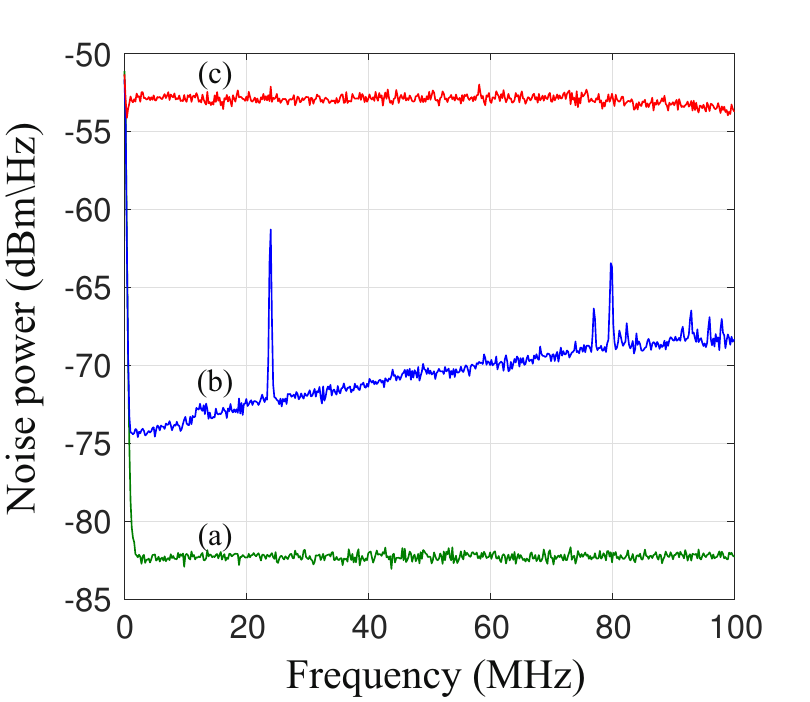}
\caption{Characteristic of homodyne detector measured with spectrum analyzer. (a) Noise of spectrum analyzer. (b) Circuit noise. (c) Shot noise. LO power is 20 mW. The data here are taken with resolution bandwidth of 200 kHz, and averaged over 300 sweeps.}
\label{fig:homodyne}
\end{figure}

To characterize the generated states, we performed homodyne measurement. The homodyne detector used here has a wide bandwidth of approximately 100 MHz with photodiodes that have the quantum efficiency higher than 98\% \cite{Serikawa2016}. The characteristic of the homodyne detector used in this experiment is shown in Fig. \ref{fig:homodyne}. Matching transversal modes between a local oscillator (LO) and measured beams is important in homodyne detection. To do so, we put the LO beam through a mode cleaning cavity which filters the transversal mode of the LO into TEM$_{00}$ and resulted in interferometric visibility of 99\% in homodyne measurement. 

A phase reference beam is introduced into the OPO for locking the phase between generated states and the LO at homodyne detector. We used two acousto-optic modulators (AOM) to shift the frequency of the phase reference beam by 100 kHz for phase locking. When the phase reference beam goes through the OPO, due to parametric amplification, the electric fields of frequency $\pm$100 kHz relative to the carrier frequency will be generated, and the resultant electric field vector will elliptically oscillate in the complex plane of the frame rotating with the carrier frequency, where the electric field will circularly rotate in the case of detuning without parametric amplification. If we look at the intensity of the amplified probe beam, the intensity oscillates at 200 kHz. By demodulating this intensity, we can lock the relative phase between phase reference beam and pump beam. Moreover, when parametrically amplified detuned phase reference beam and LO interfere the resultant intensity will oscillate with frequency of 100 kHz. We can lock the phase between probe beam and LO by properly demodulating the interference signal. This demodulation is not so trivial because the complex amplitude of phase reference beam is elliptic due to parametric amplification. By properly locking the relative phase between the phase reference beam and the LO and between the phase reference beam and the pump beam, we indirectly lock the relative phase between a cat state and the LO. Since we are locking relative phases by using demodulation of the beat signals, we can lock at any arbitrary relative phase by changing the phase of the demodulation signal. This locking method using parametrically amplified detuned probe beam gives error signal with better S/N ratio than the usual method using phase modulation because the size of the error signal is proportional to the strength of the carrier not the strength of the phase modulation.

We used Pound-Drever-Hall technique \cite{R.W.P.DreverJ.L.Hall1983} for locking the second harmonic generator. An electro-optic modulator (EOM) is put in front of the second harmonic generator for generation of the error signal. For the OPO and filtering cavities, we used tilt locking technique \cite{Shaddock1999}. We also shifted the frequency of the cavity lock beams by 200 kHz to prevent unwanted interference with other beams.

In this experiment, the phase reference beam and cavity lock beams (collectively called control beams) are turned on and off periodically by AOMs. Beside two AOMs in the path of phase reference beam and cavity lock beams, there are two more AOMs that are common to both beams in order to increase the extinction rate of the control beams. All the aforementioned cavity locking and phase locking is performed by electrical feedback using servo amplifiers when the control beams are turned on. When the control beams are turned off, the optical system is held in the same state before the control beams are turned off. Using this periodic switching, we prevented the control beams from reaching the APD, which will result in fake clicks that degrade the purity of the cat states. 

\section{Experimental results}
For the rest of the paper, we refer to the measurement where we digitally integrate the electric signal with the temporal mode as post processing, as opposed to real-time measurement where the electric signal is continuously integrated with the temporal mode by a low-pass filter (LPF) \cite{Ogawa2016}.

In this paper, we performed photon subtraction from squeezed vacua with three different squeezing level, and generated three types of quantum states: a single photon state, a cat state, and a squeezed single photon state. To characterize each state, we performed homodyne measurement for 37 phases from 0 degrees to 180 degrees by 5 degree step and we collected data of 10,000 events for each phase. The electric signal from a homodyne detector is split into two paths and we put a LPF for real-time measurement in one of the paths and simultaneously record unfiltered and filtered electric signal using an oscilloscope. To extract the quadrature values from CW homodyne measurement, we need to integrate the electric signals  from homodyne detector with the temporal mode \cite{Molmer2006}. Also, in order to design and make a LPF for real-time measurement, we need to know the shape of the temporal mode of cat states beforehand. This means that it is important that we estimate the temporal mode correctly. Using the obtained quadrature values, we estimate the density matrices and Wigner functions for both measurement using maximum likelihood method \cite{Lvovsky2004,Lvovsky2009}. 

In this section, we show the experimental results in the following order. First, we show the estimation of the temporal mode of the generated states. Then, we show the Wigner functions estimated from the quadrature values which are obtained by using the estimated temporal mode. We can evaluate the quality of the states generated via photon subtraction by the value at the origin of the Wigner functions, which we call Wigner negativity. In the ideal case, the Wigner negativity is $W(0,0)=-1/\pi\approx-0.318$ which is due to the fact that the states generated by subtracting a single photon from squeezed vacuum have only odd photon numbers \cite{Dakna1997a}. We also evaluate whether the effects of the experimental imperfections are consistent with degradation of the Wigner negativities of the estimated Wigner functions. Finally, to verify the success of the real-time measurement, we show the correlation plot between the quadrature values of post processing and real-time measurement. As a qualitative indicator, we also show the screen capture of oscilloscope recording the electric signals from homodyne detector.

First, let us look at the temporal mode of the generated states. The temporal mode of the states generated by photon subtraction depends on the response functions of the OPO and filtering cavities. For the OPO, the temporal mode localized around time $t_{0}$ is derived from the time correlation function and has the following form \cite{Collett1984};
\begin{equation}
f_{\textrm{OPO}}(t;t_{0})=\sqrt{\gamma_{\textrm{OPO}}}\exp(-\gamma_{\textrm{OPO}}\vert t-t_{0}\vert),
\label{eq:OPO}
\end{equation}
where $\gamma_{\textrm{OPO}}$ corresponds to bandwidth of the OPO, and can be written as following;
\begin{equation}
\gamma_{\textrm{OPO}}=2\pi f_{\textrm{HWHM}},
\end{equation} 

where, $f_{\textrm{HWHM}}$ is half-width half-maximum of the OPO. For filtering cavities, the response function is equivalent to an ideal Lorentzian filter and can be expressed as following in time domain \cite{Collett1984};
\begin{equation}
f_{\textrm{filter}}(t;t_{0})=\sqrt{2\gamma_{\textrm{filter}}}\exp(-\gamma_{\textrm{filter}}\vert t-t_{0}\vert)\Theta(t_{0}-t),
\label{eq:filter}
\end{equation}
where $\Theta(t)$ is Heaviside step function, and $\gamma_{\textrm{filter}}$ corresponds to the bandwidth of filtering cavities.

In our case, we have three filtering cavities. Therefore, the temporal mode of the generated states will become a time convolution between the response of the OPO in expressed by Eq. \eqref{eq:OPO} and three filtering cavities expressed by Eq. \eqref{eq:filter} with $\gamma_{\textrm{filter}}$ which corresponds to parameter of each cavity. After performing time convolution of response of the OPO and three filtering cavities, the ideal temporal mode $f_{\textrm{ideal}}(t;t_{0})$ can be expressed as following;
\begin{equation}
f_{\textrm{ideal}}(t;t_{0})=N\left[\sum_{i=1}^{4}c_{i}\exp(-\gamma_{i}\vert t-t_{0}\vert)\Theta(t_{0}-t)+\left(\sum_{i=1}^{4}c_{i}\right)\exp(-\gamma_{4}\vert t-t_{0}\vert)\Theta(t-t_{0})\right].
\label{eq:temporalmode}
\end{equation}
$N$ is a normalization constant, $i=1-3$ corresponds to filtering cavities, $i=4$ corresponds to the OPO, and $\gamma_{i}$ corresponds to bandwidth of each cavity. $c_{1}=2\gamma_{4}(\gamma_{3}-\gamma_{2})/(\gamma_{4}^{2}-\gamma_{1}^{2})$ and $c_{2},$ and $c_{3}$ are the cyclic permutation. The expression for $c_{4}$ and $N$ are
\begin{align}
c_{4}&=\frac{\gamma_{1}-\gamma_{2}}{\gamma_{4}-\gamma_{3}}+\frac{\gamma_{2}-\gamma_{3}}{\gamma_{4}-\gamma_{1}}+\frac{\gamma_{3}-\gamma_{1}}{\gamma_{4}-\gamma_{2}},\\
N&=\left[\sum_{i,j=1}^{4}\frac{c_{i}c_{j}}{\gamma_{i}+\gamma_{j}}+\frac{1}{2\gamma_{4}}\left(\sum_{i=1}^{4}c_{i}\right)^{2}\right]^{-\frac{1}{2}}.
\end{align}

From Eq. \eqref{eq:temporalmode} it is clear that $f_{\textrm{ideal}}(t;t_{0})$ is not a perfect exponentially rising temporal mode. However, when one of the $\gamma_{i}$ of the filtering cavities is much smaller than that of the OPO, which corresponds to the case where there is a narrowband filtering cavity, that term becomes dominant in Eq. \eqref{eq:temporalmode} and the temporal mode becomes close to exponentially rising.
\begin{figure}[t]
\centering\includegraphics[width=0.6\textwidth]{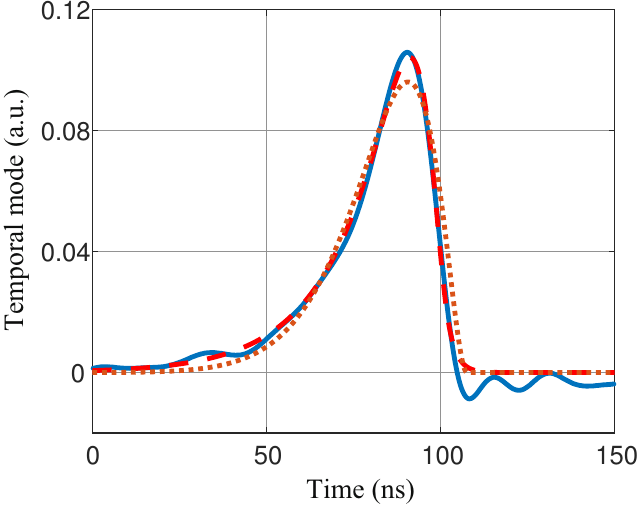}
\caption{Temporal modes of generated cat state. Solid curve: estimated temporal mode. Dashed curve: Theoretical prediction from experiment parameters. Dotted curve: Temporal response of low-pass filter used in real-time measurement.}
\label{fig:mode}
\end{figure}

We used independent component analysis (ICA) \cite{Comon1994}, where the non-Gaussianity induced via photon subtraction is utilized to find a set of independent modes and estimate the temporal modes of the generated states. We measured the quadrature for the phase that corresponds to largest variance and used the quadrature data of 10,000 events at that phase in the temporal mode estimations. Another method similar to ICA is the estimation via principal component analysis (PCA) \cite{Morin2013,MacRae2012}, which make use of the fact that the variances of the heralded states are larger than the initial squeezed vacua.

The temporal mode of the cat state is shown in Fig. \ref{fig:mode}. The solid curve is the estimated temporal mode and the dashed curve is the theoretical prediction from the experimental parameters using Eq. \eqref{eq:temporalmode}. The inner product between these two curves is 0.996 . The dotted curve is the temporal response of the 3rd-order LPF that is designed for real-time measurement. The inner product of this to the estimated temporal mode is approximately 0.988 . From these results, it is clear that the temporal mode of the generated states is consistent with the theoretical prediction and the designed LPF is also consistent with the temporal mode of the generated states. We integrated the electric signal from homodyne detector with this estimated temporal mode to obtain the quadrature values for post processing. For real-time measurements, we used the LPF which is designed to have response as in Fig. \ref{fig:mode}.

\begin{figure}[t!]
\centering\includegraphics[width=\textwidth]{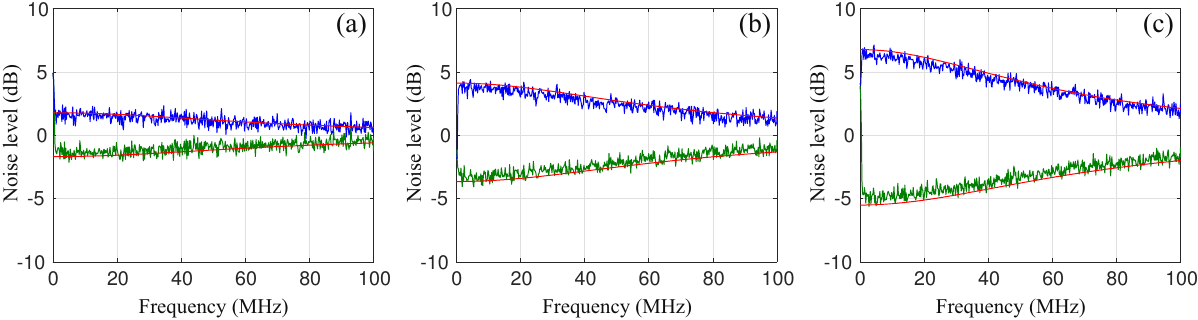}
\caption{Squeezing spectra of the initial squeezed vacua normalized to vacuum. (a) $\xi=0.11$, (b) $\xi=0.25$, and (c) $\xi=0.39$. Solid lines: theoretical plots of squeezing spectra using Eq. \eqref{eq:squeeze}. The external loss for all cases are $L=0.113$. Note that $L$ included 3\% loss due to $R=0.97$ beamsplitter used in photon subtraction.}
\label{fig:squeeze}
\end{figure}

Figure \ref{fig:squeeze} shows the squeezing spectra of the initial squeezed vacua used for generation of the states. The squeezing spectra are described by the following equation \cite{Collett1984};
\begin{equation}
S_{\pm}(f) = 1\pm(1-L)\frac{4\xi}{(1\mp\xi)^{2}+(f/f_{\textrm{HWHM}})^{2}},
\label{eq:squeeze}
\end{equation}
where $\xi$ is the normalized pump power, $L$ is total external loss, and $f_{\textrm{HWHM}}$ is the bandwidth of the OPO. In this experiment, we used the squeezed vacua with $\xi=$0.11, 0.25, 0.39 for generation of quantum states with high fidelity to single photon state, cat state, and squeezed single photon state, respectively. The solid lines in Fig. \ref{fig:squeeze} show the theoretical plot of squeezing spectra, and the losses in this experiment are as follows: 1.2\% propagation loss, 2\% loss due to quantum efficiency of photodiodes, 1\% loss due to circuit noise of the homodyne detector, 2\% loss due to visibility at homodyne measurement, and 2.1\% loss due to the escape efficiency of the OPO. Therefore, the total experimental loss in this experiment is 8.3\%. In the measurement of the squeezing spectra there is an extra 3\% loss due to $R=0.97$ used in photon subtraction. Note, however, that we ignored the frequency dependence of the circuit noise of the homodyne detector. This is because the cat states generated in this experiment has bandwidth of approximately 10 MHz, which is much narrower than the bandwidth of the homodyne detector. 

\begin{figure}[ht!]
\centering\includegraphics[width=\textwidth]{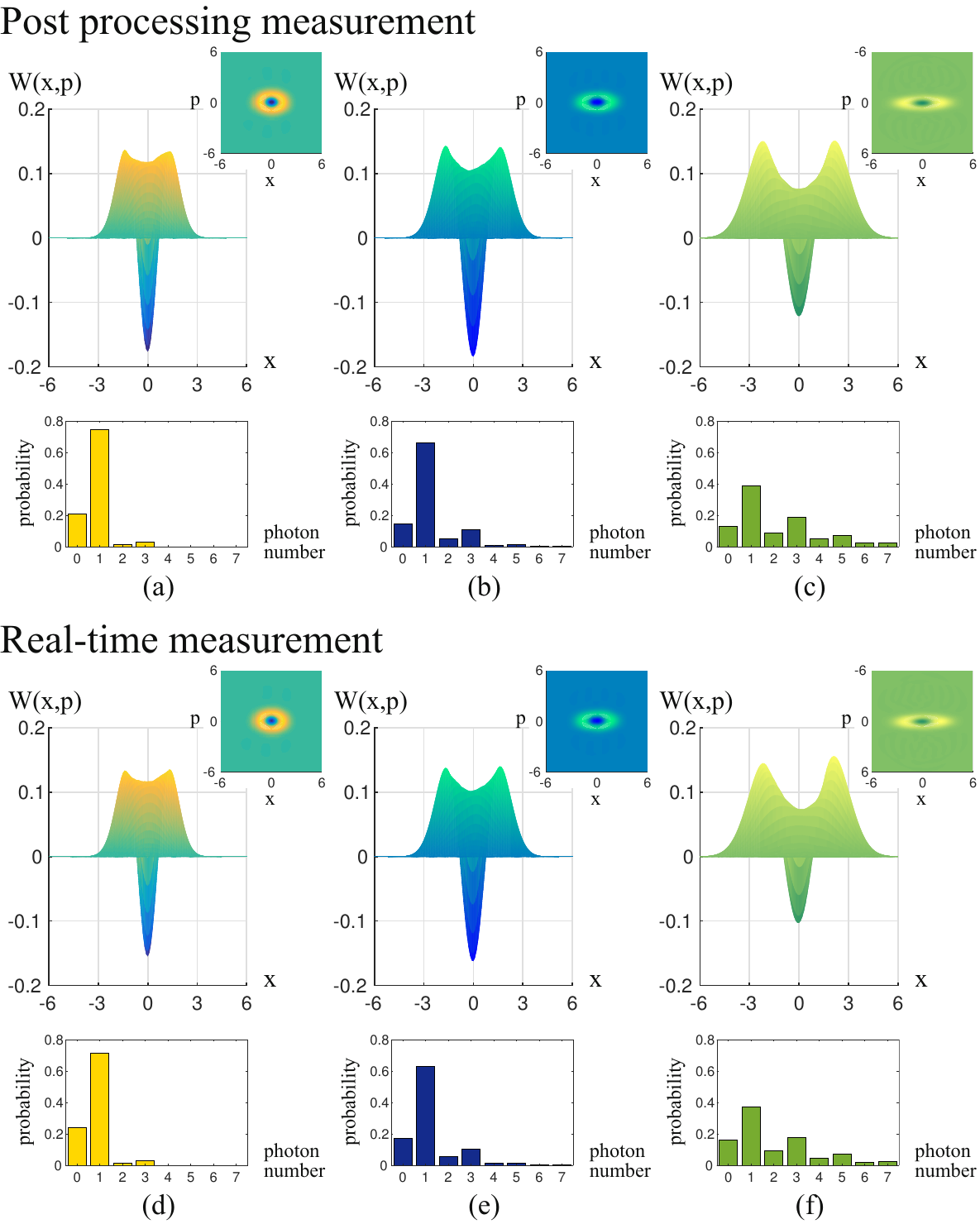}
\caption{Wigner functions and photon number probability distributions of states generated in this experiment. The upper-half is the results of post processing measurement, while the lower half is the corresponding real-time measurement. (a,d) $\xi=0.11$. (b,e) $\xi=0.25$. (c,f) $\xi=0.39$. For post processing measurements, Wigner negativities are $-0.176$, $-0.184$, and $-0.121$ respectively. For real-time measurement, Wigner negativities are $-0.154$, $-0.162$, and $-0.102$ respectively. The initial squeezing spectra are shown in Fig. \ref{fig:squeeze}.}
\label{fig:experimentresults}
\end{figure}

Figure \ref{fig:experimentresults} shows the Wigner functions and photon number probability distributions for post processing and corresponding real-time measurements.  

Figure \ref{fig:experimentresults}(a) shows the post processing measurement results of photon subtraction from squeezed vacua with $\xi = 0.11$, which corresponds to pump power of 5 mW, and initial squeezing level of 1.0 dB near DC component. The generation rate of this state is about 1,200 counts per second (cps). Because the squeezing level is low in this case, the generated state possesses high fidelity to single photon states. The ratio of single photon in the generated state is 0.74, and the Wigner negativity of the Wigner function is $-0.176$ . However, we can see from the Wigner function that this state is slightly squeezed. This means that despite the fact that the generated state consists of mostly single photon component, the generated state is not perfectly single photon and this slight squeezing is due to the initial squeezing level. Therefore, we need to lower the squeezing level even more to generate pure single photon with this method, which will result in a lower generation rate and the effect of fake counts will become more apparent.
 
Figure \ref{fig:experimentresults}(b) shows the post processing measurement results of photon subtraction from squeezed vacua with $\xi = 0.25$, which corresponds to pump power of 25 mW, and initial squeezing level of 3.0 dB near DC component for generation of quantum state with high fidelity to minus Schr{\"o}dinger's cat state $\vert\Psi_{\textrm{cat},-}(\alpha)\rangle\propto\vert\alpha\rangle-\vert-\alpha\rangle$ with $\vert\alpha\vert^{2}=1$ after photon subtraction \cite{Lund2004}. The generation rate of this state is about 4,800 cps. The Wigner function of our cat state possesses the Wigner negativity of $-0.184$ without loss corrections which is the highest value ever observed without loss correction. The fidelity $F$ to the cat state is calculated by $F=\langle\Psi_{\textrm{cat},-}(\alpha)\vert\hat{\rho}\vert\Psi_{\textrm{cat},-}(\alpha)\rangle$, where $\hat{\rho}$ is the density matrix of our state. The resulting fidelity is $F=0.782$ to Schr{\"o}dinger's cat state with $\vert\alpha\vert^{2}=1.02$, which is also the highest value ever observed without loss correction. Since a Schr{\"o}dinger's cat state possesses only odd photon numbers, the highest fidelity possible is equal to the sum of the odd photon numbers probability, which in our case is 78.8\%. This means that the odd photon subspace of this state has very high fidelity to Schr{\"o}dinger's cat state.

Figure \ref{fig:experimentresults}(c) shows the post processing measurement results of photon subtraction from squeezed vacua with $\xi = 0.39$, which corresponds to pump power of 60 mW, and initial squeezing level of 5.0 dB near DC component. The generation rate is about 16,000 cps. The Wigner function of squeezed single photon state has the Wigner negativity of $-0.121$, which means that this state also possess high nonclassicality. The degradation of the Wigner negativity of this state with $x=0.39$ is more than that of states with $x=0.11,0.25$. This is due to the fact that the state with $x=0.39$  has more mean photon number than the other two states which we will address this in more details. In \cite{Wakui2007a} the photon subtraction is taken out with initial squeezing level of 3.7 dB as a demonstration of photon subtraction from highly squeezed vacua. In that experiment the four photon probability is larger than five photon probability. On the other hand, even with large squeezing, the odd photon probabilities are larger than the adjacent even photon probabilities in our experiment, and we can observe large photon number components clearly. We believe that this is the effect of narrowband filtering \cite{Junichi2017}, which becomes more apparent as the initial squeezing level increase.

Next, we calculate whether the experimental imperfections and the Wigner negativities of the Wigner functions are consistent with each other. From the Wigner negativity of Wigner function, we calculate the sums of the even photon number probability to be 22.4\%, 21.2\%, and 31.0\% for the $\xi=$0.11, 0.25, and 0.39, respectively. For the ideal states generated from photon subtraction, these values will be 0 and there are two reasons for the deviation from the ideal value: First, the photon losses and detection losses. Second, the imperfections of photon subtraction. The former corresponds to the contamination of the vacua in the generated states, while the latter corresponds to mixing between the generated states and initial squeezed vacua. We refer to the former as loss and the latter as mixedness to differentiate between these two types of degradation of the Wigner negativity. In case of losses, the sum of the even photon number probability will become more as the mean number photon of the generated states become higher. This is because the photon losses from Fock states scale with the number of photons, i.e. $\hat{a}\vert n\rangle=\sqrt{n}\vert n-1\rangle$, where $\vert n\rangle$ is a photon number state and $\hat{a}$ is an annihilation operator. On the other hand, the mixedness is how much the squeezed vacua is mixed with the generated states, thus the increment of the sum of the even photon number probability due to mixednesses does not depend on the mean photon number.

\begin{table}[t!]
\caption{Mixedness due to the imperfection in the photon subtraction}
\centering
\begin{tabular}{l c c c}
\hline
&\multicolumn{3}{c}{Quantum states}\\
\cline{2-4}
&Single photon &Schr{\"o}dinger's cat&Squeezed single  photon\\ 
\hline
Fake counts&2.7\%&0.7\%&0.2\%\\
Two photons subtraction&1.7\%&2.7\%&4.8\%\\
\hline
total&4.4\%&3.4\%&5.0\%\\
\hline
\end{tabular}
\label{tab:fake}
\end{table}

As mentioned before, the total loss in this experiment is 8.3\%. However, in addition to losses, there are mixednesses which arose from the imperfections of the photon subtraction. The main sources of mixedness are fake counts and two photon subtraction. Fake counts can be measured directly, and probability of two photon subtraction can be calculated using the normalized pump amplitude $x$ and reflectivity of beamsplitter $R$ used in photon subtraction \cite{Dakna1997a}. Here, we ignore the probability of subtracting more than two photons. The mixedness are shown in Table \ref{tab:fake}. These imperfections mix the generated states with states with even photon number states, which degrades the Wigner negativities.

From both losses and mixednesses, let us approximate the ratio of even photon numbers from experimental parameters. Using squeezing level (equivalently, normalized pump amplitude $\xi$) and the reflectivity of beamsplitter $R$, we can calculate the ideal states generated by photon subtraction \cite{Dakna1997a}. Then, by applying photon loss and mixing with squeezed vacua due to the mixednesses on the ideal states, we can estimate the theoretical ratio of even photons in each state from experimental parameters. For the simplicity of the estimation, we assumed that one photon losses are dominant, and ignore multiple photon losses. 
Calculating this for each states, the estimation of the sum of even photon number probabilities are approximately 14\%, 16\%, and 23\%, while the experimental results are 22.4\%. 21.2\%, and 31.0\%, respectively. The value of the estimation and the experimental results are in agreement to each other, and the difference might come from parameters that we cannot measure directly or accurately such as quantum efficiency of photodiodes, or other factors we cannot take into account such as accuracy of state tomography. Also, the theoretical plots of squeezing spectra in Fig. \ref{fig:squeeze}, though in good agreement with measurement results, have less losses than the experimentally measured spectra. This indicates that the estimated losses are too little, which explains the difference in estimation of even photon number. 

\begin{figure}[htp!]
\centering\includegraphics[width=\textwidth]{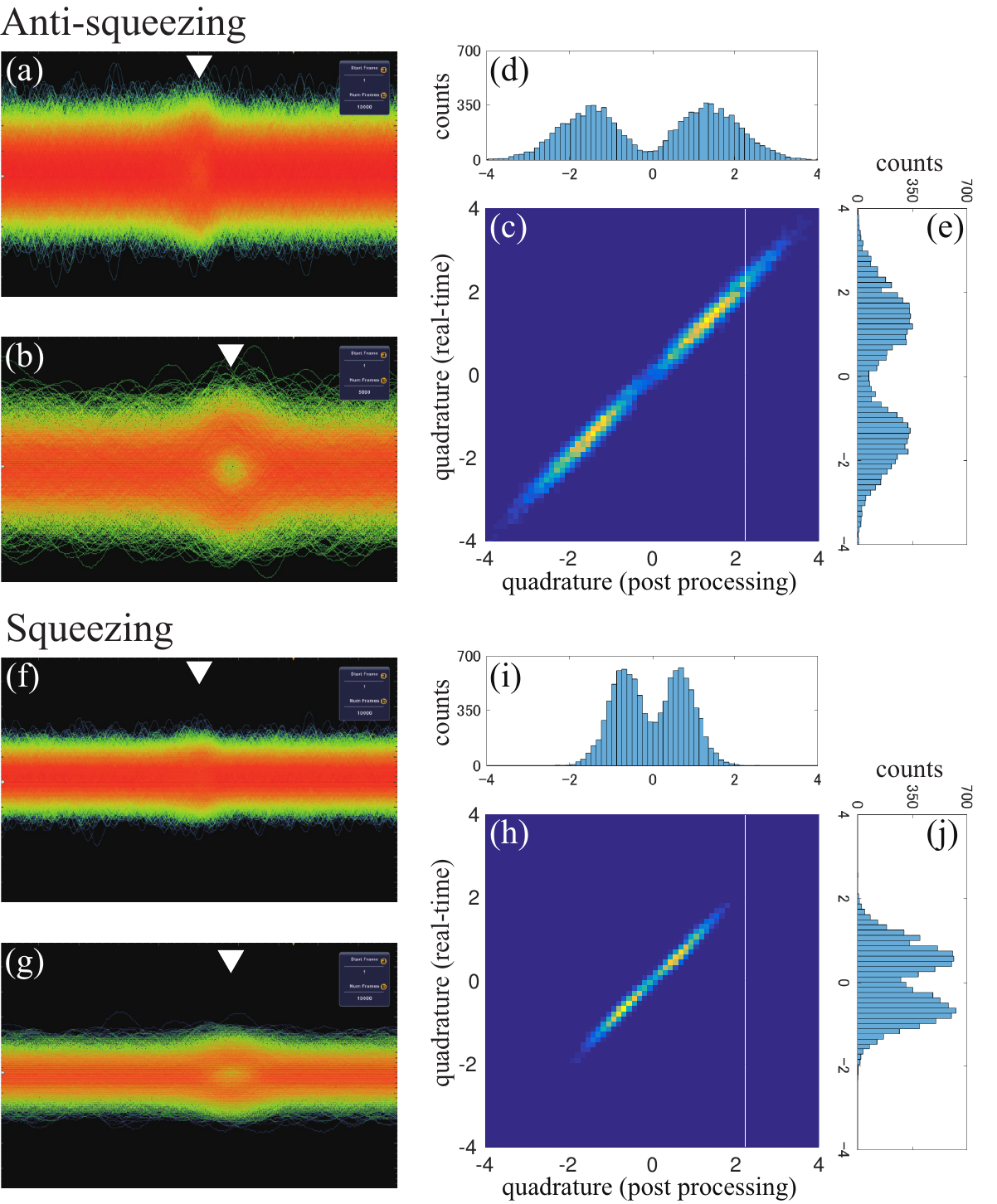}
\caption{Screen captures of oscilloscope displaying electric signal from homodyne detector and quadrature distribution of cat state of phases corresponded to anti-squeezing and squeezing. The number of the overlaid events is 10,000 and electric signal is recorded 200 ns around the timing of photon detection. For anti-squeezing: (a) Screen capture of post processing measurement. (b) Screen capture of real-time measurement. White triangle marks represent the timing of the generation of cat states ($\xi=0.25$). The coloring represents frequency of the distribution at each time and change from blue to green, yellow and red as the frequency increase. (c) Correlation plot between quadrature of post processing and real-time measurement. (d) Quadrature distribution of post processing. (e) Quadrature distribution of real-time measurement (This corresponds to the histogram of electric signal at the white triangle mark in (b)) . The correspoding figures for squeezing are (f-j). (See \textcolor{blue}{Visualization 1} for the screen captures of real-time measurement for phases in between.)}
\label{fig:realtime}
\end{figure}

Figures \ref{fig:experimentresults}(d), \ref{fig:experimentresults}(e), and \ref{fig:experimentresults}(f) show the results of real-time measurement in the case of $\xi=$0.11, 0.25, 0.39, respectively. Qualitatively, the Wigner functions is in good agreement with results of post processing. The Wigner negativity of each state is $-0.154$, $-0.162$, and $-0.102$ . Although the Wigner negativity of real-time measurement is still high, it is smaller than the corresponding post processing results. However, the difference in the Wigner negativity, which is approximately $-0.02$, is the same for all states. This indicates that the degradation of the Wigner negativities should be caused by common sources. When the electric signal from the homodyne detector is integrated with the mode that is not perfectly matched to the temporal mode of the generated states, the unmatched portion will pick up squeezed vacua instead of the state generated from photon subtraction. In this experiment, mode mismatch between the LPF and the temporal mode corresponded to $1-0.988^{2}=2.4\%$ extra mixedness in real-time measurements, which increase the Wigner negativity by 0.015 from post-processing results. Thus, mode mismatch between LPF and temporal mode adds an additional mixedness in real-time measurement.

As mentioned before, we can qualitatively verify the success of the real-time measurement by the using screen capture of the oscilloscope, and quantitatively by correlation plots of the quadrature values between the post processing and the real-time measurements. Figures \ref{fig:realtime}(a) and \ref{fig:realtime}(b) show the screen captures of oscilloscope for post-processing and real-time measurements of cat state ($\xi=0.25$) at the phase of anti-squeezing, and Fig. \ref{fig:realtime}(f) and \ref{fig:realtime}(g) show the corresponding at the phase of squeezing. In the case of post processing, the histogram of electrical signal does not possess the characteristics of cat states at any arbitrary timing. In the case of the real-time measurement, if we look at the electric signal at the timing corresponding to the generation of cat states, at the triangle mark, the histogram of electric signal possess the characteristics of quadrature distribution of cat states; a dip around quadrature equals to 0 and two peaks, and the variance for anti-squeezing and squeezing are different. For the other phases, we also got the histogram of the electric signals which possess the characteristics of the cat state at the same timing (See \textcolor{blue}{Visualization 1}), which is the qualitative indicator that we succeeded in the real-time quadrature measurement. In addition, the screen captures of the oscilloscope are symmetric with respect to the generation timing in real-time measurement. This symmetry is another qualitative indicator of the success of real-time measurement \cite{Ogawa2016}. 

If the real-time measurement is successfully performed, the quadrature values obtained by real-time measurement and post processing measurement will have same values. Figure \ref{fig:realtime}(c) and (h) shows the correlation between the quadrature of post processing and real-time measurements at the phase corresponded to anti-squeezing and squeezing, respectively. Figures \ref{fig:realtime}(d,i) and \ref{fig:realtime}(e,j) show the quadrature distribution of post-processing and real-time measurements, respectively. From Fig. \ref{fig:realtime}(c) and (h), we can see that the quadrature values of both measurements are highly correlated for both anti-squeezing and squeezing. The correlation plots for all the other phases also show the similar correlations, and the correlation coefficients between post processing and real-time measurements are over 0.99 for all phases. 

\section{Conclusion}
We generated highly pure Schr{\"o}dinger's cat states with a new method via optical filtering. It has been shown that this method has no theoretical limit on the purity of generated states as opposed to the previous photon subtraction. We succeeded in the generation of Schr{\"o}dinger's cat state that has a new Wigner negativity record of $-0.184$ without any loss corrections. We also succeeded in the generation of single photon-like state and squeezed single photon state with this method. From another aspect, optical filtering can be considered as temporal mode shaping, and the temporal mode becomes exponentially rising with our method, which allows us to demonstrate a real-time quadrature measurement. In the real-time measurement, the states we generated showed high nonclassicality and the quadrature values are highly correlated with that of the post processing. This indicates that we succeeded in the real-time measurement of cat states, which is the first real-time quadrature measurement of quantum states with multiple photons and phase sensitivity.

This method can be easily extended to any quantum states that can be generated by photon subtraction or heralding, and allows us to perform photon subtraction from highly squeezed vacua. Therefore, this method is the most promising and realistic in the generation of highly pure quantum states and, at the same time, temporal mode shaping for the applications in MBQC. We expect that photon subtraction with narrowband optical filtering will become a new basis for generation of even more complex highly pure nonclassical states, which will accelerate the development of quantum information processing.

\section*{Funding}
This work was partly supported by CREST (JP- MJCR15N5) of Japan Science and Technology Agency (JST), KAKENHI of Japan Society for the Promotion of Science (JSPS), and APSA of Ministry of Education, Culture, Sports, Science and Technology of Japan (MEXT). 

\end{document}